\pdfoutput=1
\documentclass[a4paper,11pt]{article}
\usepackage{pos}
\renewcommand{\logo}{\relax} 

\title{Update on transverse-spin azimuthal asymmetries\\
in single-hadron SIDIS\\
on transversely polarised deuterons at COMPASS}
\ShortTitle{Transverse-spin asymmetries on deuterons at COMPASS}

\author*[a]{Artur Hoghmrtsyan}
\onbehalf{on behalf of the COMPASS Collaboration}

\affiliation[a]{A. Alikhanyan National Science Laboratory,\\
2 Alikhanyan brothers street, Yerevan, Armenia}

\emailAdd{artur.hoghmrtsyan@cern.ch}

\abstract{Over the years, the COMPASS experiment at CERN has pursued an
extensive SIDIS programme aimed at a comprehensive understanding of the
structure of the nucleon, using a high-energy muon beam and polarised
targets. Within this programme, measurements with a transversely polarised
deuteron target were performed in 2022 using the $160\,\mathrm{GeV}/c$ muon
beam. In single-hadron SIDIS on such a target, eight transverse-spin-dependent
azimuthal asymmetries arise, providing sensitivity to several
transverse-momentum-dependent parton distribution functions and to
subleading-twist partonic correlations. We report the simultaneous
extraction of all eight asymmetries from the complete 2022 deuteron sample.
The Collins and Sivers asymmetries, previously presented using about 70\% of
the 2022 statistics, are updated with the full sample and found consistent
with the earlier result, while the six remaining asymmetries are presented
for the first time with the full 2022 statistics. Combined with the existing
proton results, these measurements complete a comprehensive COMPASS programme
on transverse-spin effects in SIDIS, providing essential input for global
analyses.}

\FullConference{The 33rd International Workshop on Deep Inelastic Scattering and Related Subjects (DIS2026)\\
4--8 May 2026\\
Bologna, Italy\\}

\begin{document}
\maketitle
\captionsetup{hypcap=false}

\section{Introduction}

Semi-inclusive deep-inelastic scattering (SIDIS), $\mu$ N $\rightarrow \mu^\prime$ hX, in
which at least one hadron is detected in coincidence with the scattered
lepton, is a powerful tool to probe the spin structure of the nucleon. For
hadron transverse momenta small compared with the hard scale of the process,
the SIDIS cross section factorises into transverse-momentum-dependent parton
distribution functions (TMD PDFs) and fragmentation functions (FFs). The TMD
PDFs
generalise the collinear parton distributions by retaining the intrinsic
transverse momentum of the quarks, and its correlations with the nucleon and
quark spins give rise to characteristic azimuthal modulations of the
produced-hadron
distribution~\cite{Bacchetta:2007,Anselmino:2020,Avakian:2019}.

In the $\gamma^*$N frame, $\phi_{\rm h}$ and $\phi_{\rm S}$ denote the
azimuthal angles of the produced hadron and of the target spin, measured
about the virtual-photon direction relative to the lepton scattering plane
(see Fig.~\ref{fig:sidis-kinematics}).

\begin{center}
  \centering
  \begin{minipage}[c]{0.42\textwidth}
    \centering
    \includegraphics[height=3.0cm]{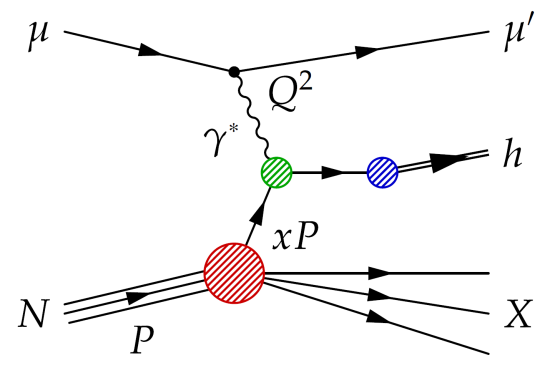}
  \end{minipage}\hfill
  \begin{minipage}[c]{0.56\textwidth}
    \centering
    \includegraphics[height=3.6cm]{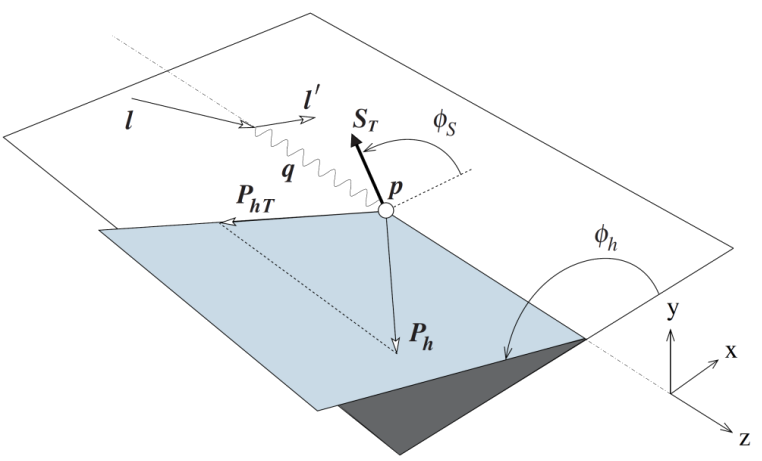}
  \end{minipage}
  \captionof{figure}{Left: the SIDIS process $\mu$ N $\rightarrow \mu^\prime$ hX. Right: the
  angles $\phi_{\rm h}$ and $\phi_{\rm S}$ in the $\gamma^*$N frame, with $l$
  ($l'$) the lepton momenta, $q$ the virtual-photon momentum, $S_{\rm T}$ the
  target transverse spin, and $P_{\rm hT}$ the transverse component of the
  hadron momentum $P_{\rm h}$, whose magnitude is denoted $p_{\rm T}$.}
  \label{fig:sidis-kinematics}
\end{center}

The part of the single-hadron SIDIS cross section that depends on the
transverse target polarisation can be written
as~\cite{Bacchetta:2007,Kotzinian:1995}
\begin{align}
\left.\frac{\mathrm{d}\sigma}{\mathrm{d}\Omega}\right|_{S_{\rm T}}
\propto S_{\rm T}\bigl\{&
 A_{\rm UT}^{\sin(\phi_{\rm h}-\phi_{\rm S})}\sin(\phi_{\rm h}-\phi_{\rm S})
 +\epsilon A_{\rm UT}^{\sin(\phi_{\rm h}+\phi_{\rm S}-\pi)}
       \sin(\phi_{\rm h}+\phi_{\rm S}-\pi) \notag\\
&+\epsilon A_{\rm UT}^{\sin(3\phi_{\rm h}-\phi_{\rm S})}\sin(3\phi_{\rm h}-\phi_{\rm S})
 +\sqrt{2\epsilon(1+\epsilon)}A_{\rm UT}^{\sin\phi_{\rm S}}\sin\phi_{\rm S} \notag\\
&+\sqrt{2\epsilon(1+\epsilon)}A_{\rm UT}^{\sin(2\phi_{\rm h}-\phi_{\rm S})}
       \sin(2\phi_{\rm h}-\phi_{\rm S})
 +\lambda\sqrt{1-\epsilon^2}A_{\rm LT}^{\cos(\phi_{\rm h}-\phi_{\rm S})}
       \cos(\phi_{\rm h}-\phi_{\rm S}) \notag\\
&+\lambda\sqrt{2\epsilon(1-\epsilon)}A_{\rm LT}^{\cos\phi_{\rm S}}\cos\phi_{\rm S}
 +\lambda\sqrt{2\epsilon(1-\epsilon)}A_{\rm LT}^{\cos(2\phi_{\rm h}-\phi_{\rm S})}
       \cos(2\phi_{\rm h}-\phi_{\rm S})\bigr\}.
\label{eq:transverse-cross-section}
\end{align}
Here $\lambda$ is the longitudinal beam polarisation and $\epsilon$ is the
ratio of longitudinal and transverse virtual-photon fluxes. Each asymmetry
amplitude in Eq.~\eqref{eq:transverse-cross-section} corresponds to a
structure function that depends on the Bjorken variable $x$, the
virtual-photon virtuality $Q^2$, the fraction $z$ of the virtual-photon
energy carried by the produced hadron, and the hadron transverse momentum
$p_{\rm T}$. The
superscript indicates the azimuthal modulation, and the two subscripts denote
the beam and target polarisation states (U unpolarised, L longitudinal,
T transverse). Five of the eight amplitudes depend only on the target
transverse spin $S_{\rm T}$ and can be measured with an unpolarised beam. They are
referred to as single-spin asymmetries. The remaining three, which enter
multiplied by $\lambda$, depend also on the longitudinal beam polarisation
and are beam--target double-spin asymmetries.

Within the TMD framework, four of the eight asymmetries have a
leading-twist interpretation~\cite{Bacchetta:2007}. The first two are the Collins
asymmetry, generated by the coupling of the transversity distribution $h_1$
to the chiral-odd Collins FF $H_1^\perp$, and the Sivers
asymmetry, which probes the Sivers function $f_{\rm 1T}^\perp$, correlating the
transverse momentum of an unpolarised quark with the transverse spin of the
nucleon. The other two are the $\sin(3\phi_{\rm h}-\phi_{\rm S})$
single-spin asymmetry, which involves the pretzelosity distribution
$h_{\rm 1T}^\perp$ coupled to the Collins FF, and the
$\cos(\phi_{\rm h}-\phi_{\rm S})$ double-spin asymmetry, usually called the
Kotzinian--Mulders asymmetry, which is sensitive to the worm-gear
distribution $g_{\rm 1T}$ convoluted with the unpolarised FF
$D_1$~\cite{Bacchetta:2007,Bastami:2019}.

The four remaining asymmetries, $A_{\rm UT}^{\sin\phi_{\rm S}}$,
$A_{\rm UT}^{\sin(2\phi_{\rm h}-\phi_{\rm S})}$, $A_{\rm LT}^{\cos\phi_{\rm S}}$, and
$A_{\rm LT}^{\cos(2\phi_{\rm h}-\phi_{\rm S})}$, receive subleading-twist contributions and
are $1/Q$-suppressed. Their structure functions can combine
several twist-2 and genuine twist-3 distribution functions and FFs,
so an individual amplitude cannot in general be identified with a single TMD
function~\cite{Bacchetta:2007,Bastami:2019,Mao:2014}. On transversely polarised protons,
COMPASS and HERMES measured predominantly small amplitudes compatible with
zero for these four asymmetries~\cite{COMPASS:SixTSA2017,HERMES:2020}.

Precise deuteron measurements provide a different flavour weighting than the
proton data and are therefore essential for separating the up- and
down-quark contributions. The limited precision of the earlier deuteron
data~\cite{COMPASS:Deuteron2007,COMPASS:Deuteron2009} restricted the
determination of the down-quark transversity and Sivers distributions in
combined proton--deuteron extractions~\cite{Martin:2015,Martin:2017}. This
motivated the high-statistics COMPASS measurement performed in 2022 with a
transversely polarised deuteron target~\cite{COMPASS:Deuteron2024}. The
measurements were performed with a $160\,\mathrm{GeV}/c$ longitudinally
polarised $\mu^+$ beam, with a polarisation of approximately $-80\%$, and an
average target polarisation above $40\%$ but well below $50\%$.
In this paper we present the results of the simultaneous extraction of all
eight transverse-spin asymmetries from the complete 2022 single-hadron
sample. The
updated full-statistics Collins and Sivers results and the first
full-statistics results for the six additional modulations are presented
below.

\section{Results}

The data selection and the analysis procedure applied for the extraction of
the eight asymmetries from the complete 2022 deuteron sample are identical to
those used for the published Collins and Sivers analysis based on
approximately 70\% of the statistics~\cite{COMPASS:Deuteron2024}. The
standard DIS selection requires $Q^2>1\,(\mathrm{GeV}/c)^2$,
$W>5\,\mathrm{GeV}/c^2$, $0.1<y<0.9$, and $0.003<x<0.7$; charged hadrons
are required to have $z>0.2$ and $p_{\rm T}>0.1\,\mathrm{GeV}/c$. After all
selections, the sample contains approximately 114 million charged hadrons.
All eight asymmetries are extracted simultaneously in a two-dimensional
$(\phi_{\rm h},\phi_{\rm S})$ extended unbinned maximum-likelihood fit, and the
amplitudes are corrected by the target polarisation, the dilution factor,
and the depolarisation factors. A detailed description of the COMPASS spectrometer
can be found in Ref.~\cite{COMPASS:Spectrometer2007}. All the COMPASS
results presented below are preliminary.

\subsection{Collins and Sivers asymmetries}

The Collins asymmetry gives access to the transversity distribution $h_1$,
which describes transversely polarised quarks in a transversely polarised
nucleon, while the Sivers asymmetry gives access to the Sivers function
$f_{\rm 1T}^\perp$, which describes the correlation between the intrinsic
transverse momentum of unpolarised quarks and the transverse spin of the
nucleon~\cite{Bacchetta:2007,Anselmino:2020}. The full-statistics 2022
deuteron measurements and the results obtained in 2010 on the proton target
are shown in Fig.~\ref{fig:collins-sivers}.

\begin{center}
  \centering
  \begin{minipage}[t]{0.49\textwidth}
    \centering
    \includegraphics[width=\linewidth]{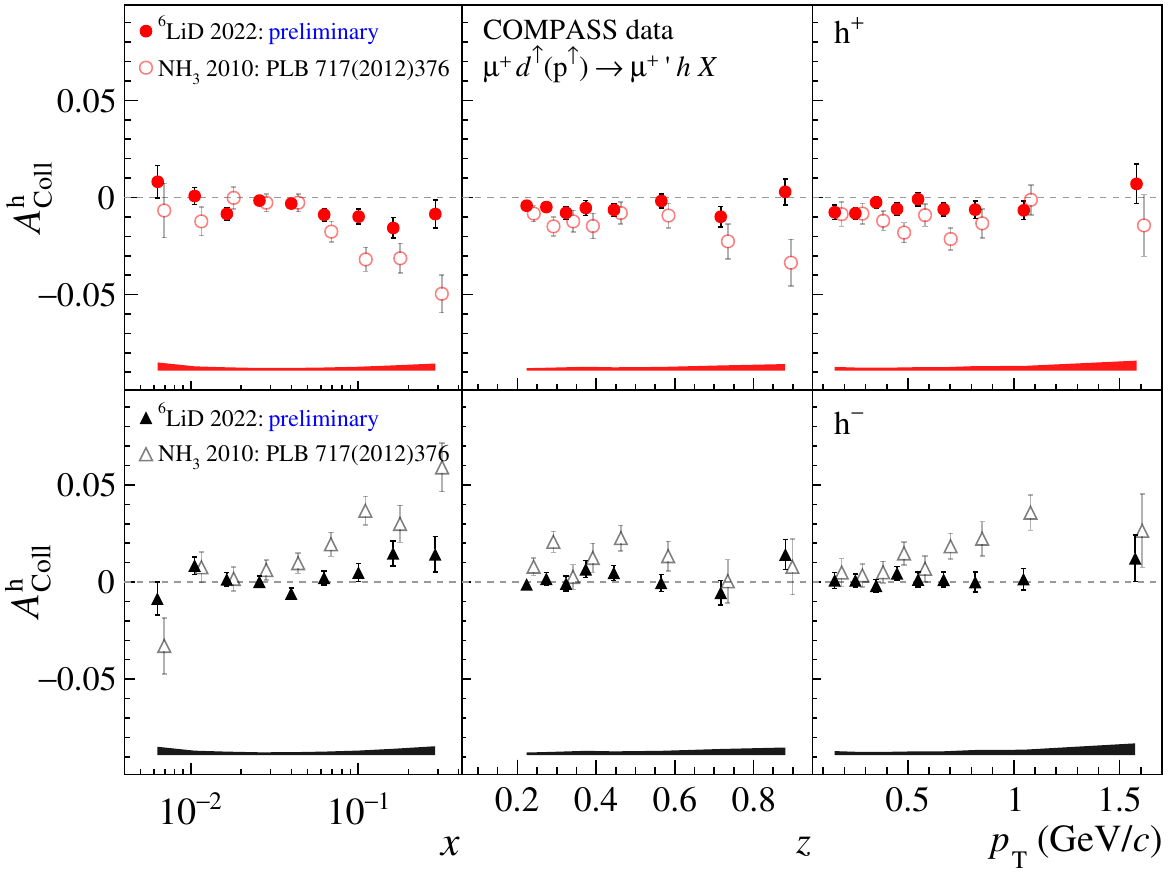}
  \end{minipage}\hfill
  \begin{minipage}[t]{0.49\textwidth}
    \centering
    \includegraphics[width=\linewidth]{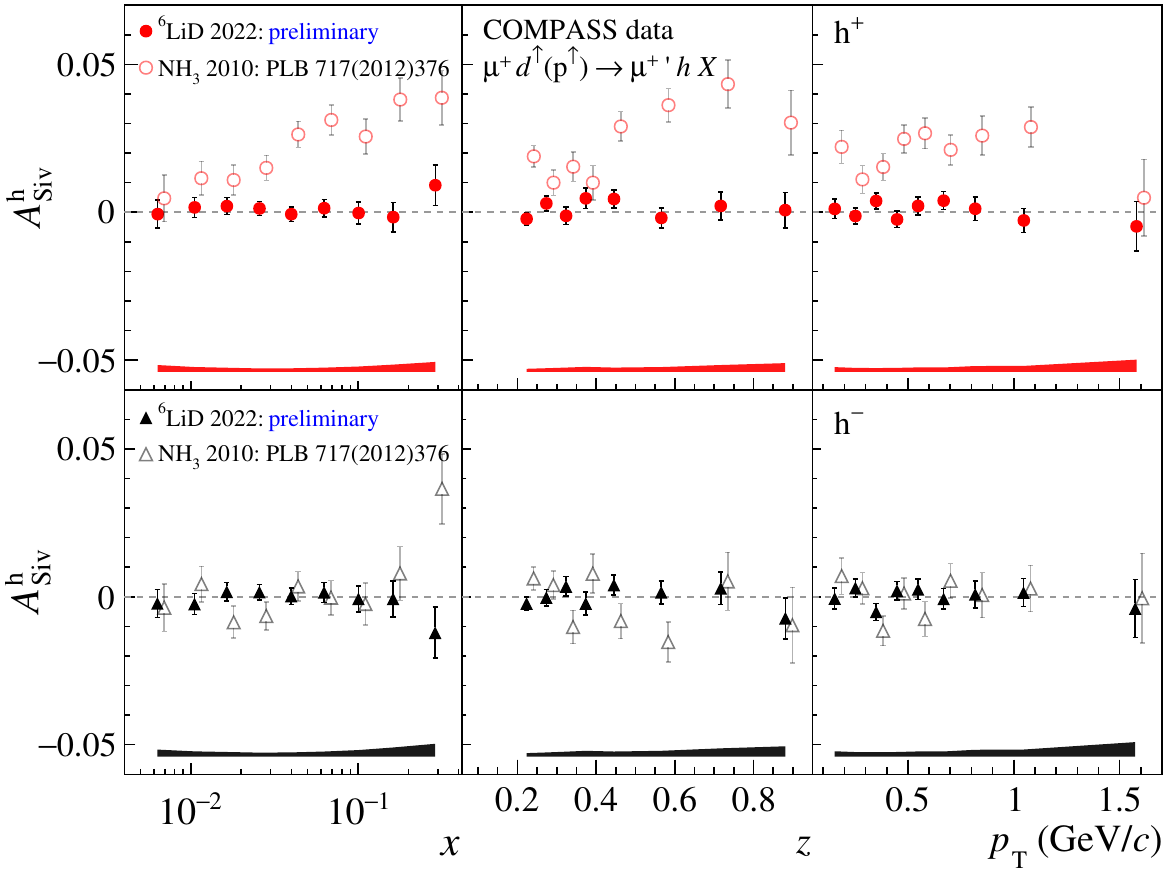}
  \end{minipage}
  \captionof{figure}{Full-statistics 2022 deuteron Collins (left) and Sivers
  (right) asymmetries as functions of $x$, $z$, and $p_{\rm T}$ for positive (top,
  red circles) and negative (bottom, black triangles) hadrons. The filled
  markers show the 2022 deuteron results (COMPASS preliminary), and the open
  markers the COMPASS 2010 proton
  results~\cite{COMPASS:Collins2012,COMPASS:Sivers2012}. The error bars
  represent the statistical uncertainties; the bands mark the systematic
  uncertainties of the 2022 results.}
  \label{fig:collins-sivers}
\end{center}

Across $x$, $z$, and $p_{\rm T}$, the complete-sample results are consistent with
the previously published results based on approximately 70\% of the 2022
statistics~\cite{COMPASS:Deuteron2024}. The deuteron Collins
amplitudes are generally small. In the largest-$x$ region they show a tendency
towards opposite signs for positive and negative hadrons, while no similarly
clear pattern is visible in the $z$ and $p_{\rm T}$ projections. The deuteron Sivers
amplitudes remain small and compatible with zero within uncertainties over the
measured $x$, $z$, and $p_{\rm T}$ ranges. Compared with the results obtained
in 2010 on the proton target, shown in the same figure, the deuteron
amplitudes are substantially smaller,
most notably for the Sivers modulation
~\cite{COMPASS:Collins2012,COMPASS:Sivers2012}.

The improved deuteron precision is particularly important for flavour
separation. In combination with proton measurements, it can strengthen the
constraints on the down-quark transversity
distribution~\cite{Martin:2015,COMPASS:Deuteron2024}.

\subsection{Other leading-twist asymmetries}

The structure function of the $\sin(3\phi_{\rm h}-\phi_{\rm S})$ pretzelosity asymmetry
contains an additional quadratic transverse-momentum weight compared with the
Collins and Sivers terms, and phenomenological studies consequently expect a
small asymmetry~\cite{Bacchetta:2007,Bastami:2019}. The first measurement with the complete
2022 deuteron sample gives small amplitudes compatible with zero throughout
the measured range, with a precision comparable to the previous COMPASS and
HERMES proton measurements~\cite{COMPASS:SixTSA2017,HERMES:2020}.

For the Kotzinian--Mulders asymmetry, the full-statistics deuteron
amplitudes are compatible with zero within uncertainties. Earlier proton
measurements showed larger, predominantly positive amplitudes
~\cite{COMPASS:SixTSA2017,HERMES:2020}. In the deuteron $x$ projection, both
hadron charges show a possible positive tendency at large $x$, but the
uncertainties do not support a claim of a nonzero signal. These two
full-statistics results are displayed in Fig.~\ref{fig:other-leading-twist}.

\begin{center}
  \centering
  \begin{minipage}[t]{0.49\textwidth}
    \centering
    \includegraphics[width=\linewidth]{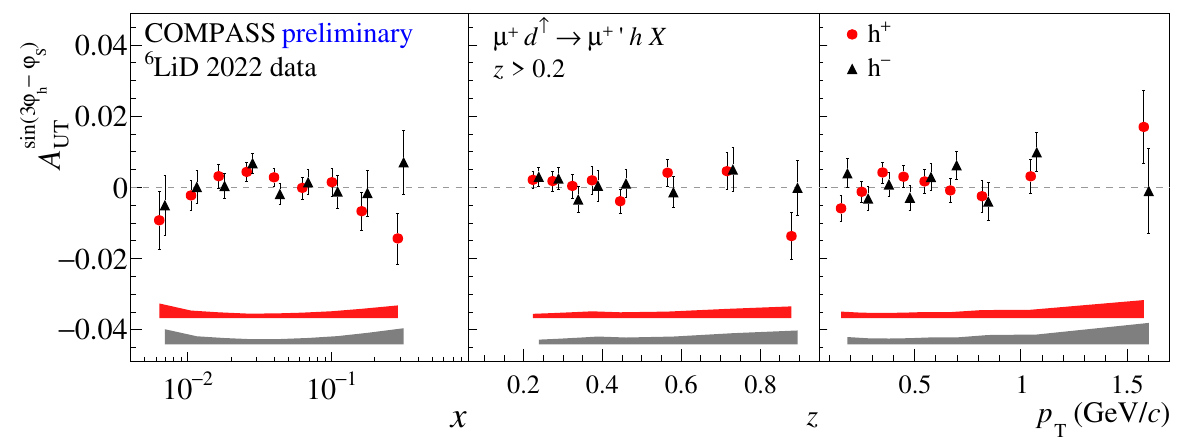}
  \end{minipage}\hfill
  \begin{minipage}[t]{0.49\textwidth}
    \centering
    \includegraphics[width=\linewidth]{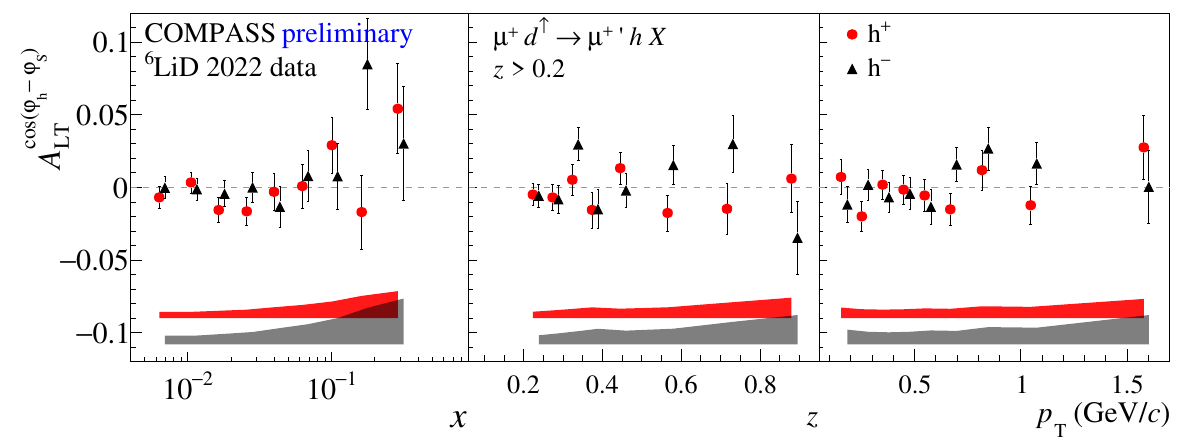}
  \end{minipage}
  \captionof{figure}{Full-statistics 2022 deuteron results for the
  $\sin(3\phi_{\rm h}-\phi_{\rm S})$ pretzelosity (left) and
  $\cos(\phi_{\rm h}-\phi_{\rm S})$ Kotzinian--Mulders (right) asymmetries as functions of
  $x$, $z$, and $p_{\rm T}$ for positive (red circles) and negative (black
  triangles) hadrons. The error bars represent the statistical uncertainties;
  the red and grey bands mark the systematic uncertainties for positive and
  negative hadrons, respectively. The results are COMPASS preliminary.}
  \label{fig:other-leading-twist}
\end{center}

Across the $x$, $z$, and $p_{\rm T}$ projections, the pretzelosity amplitudes fluctuate around
zero without a coherent kinematic dependence, whereas the
Kotzinian--Mulders amplitudes show somewhat larger fluctuations and the
large-$x$ tendency noted above. No comparable systematic pattern is visible
in their $z$ or $p_{\rm T}$ dependence. Being sensitive to a different quark-flavour
combination than the proton data, these measurements provide complementary
input for global extractions of $h_{\rm 1T}^{\perp}$ and $g_{\rm 1T}$. 

\subsection{Subleading-twist asymmetries}

In addition to the $1/Q$ factor, the four subleading-twist asymmetries are
subject to further kinematic
suppressions~\cite{Bacchetta:2007,Bastami:2019,Mao:2014}.
Figure~\ref{fig:subleading-twist} presents the four asymmetries from the
complete 2022 deuteron sample as functions of $x$, $z$, and $p_{\rm T}$. The
amplitudes are generally small and compatible with zero within uncertainties.
In particular, the measured $A_{\rm UT}^{\sin\phi_{\rm S}}$ amplitudes remain below
1\%. The $A_{\rm UT}^{\sin(2\phi_{\rm h}-\phi_{\rm S})}$ and
$A_{\rm LT}^{\cos(2\phi_{\rm h}-\phi_{\rm S})}$ results show no coherent kinematic trend,
while $A_{\rm LT}^{\cos\phi_{\rm S}}$ exhibits larger point-to-point fluctuations but no
stable deviation from zero. 

\begin{center}
  \centering
  \begin{minipage}[t]{0.49\textwidth}
    \centering
    \includegraphics[width=\linewidth]{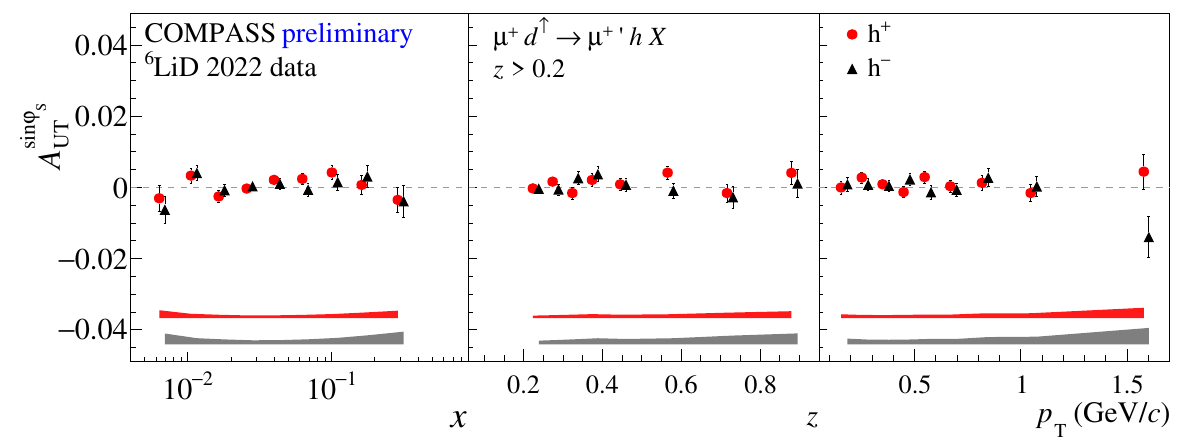}
  \end{minipage}\hfill
  \begin{minipage}[t]{0.49\textwidth}
    \centering
    \includegraphics[width=\linewidth]{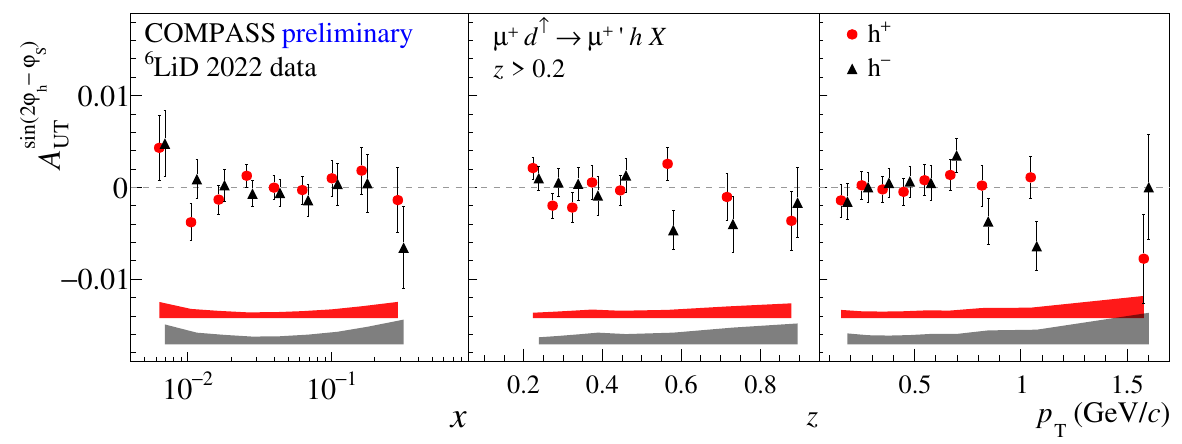}
  \end{minipage}\\[0.5em]
  \begin{minipage}[t]{0.49\textwidth}
    \centering
    \includegraphics[width=\linewidth]{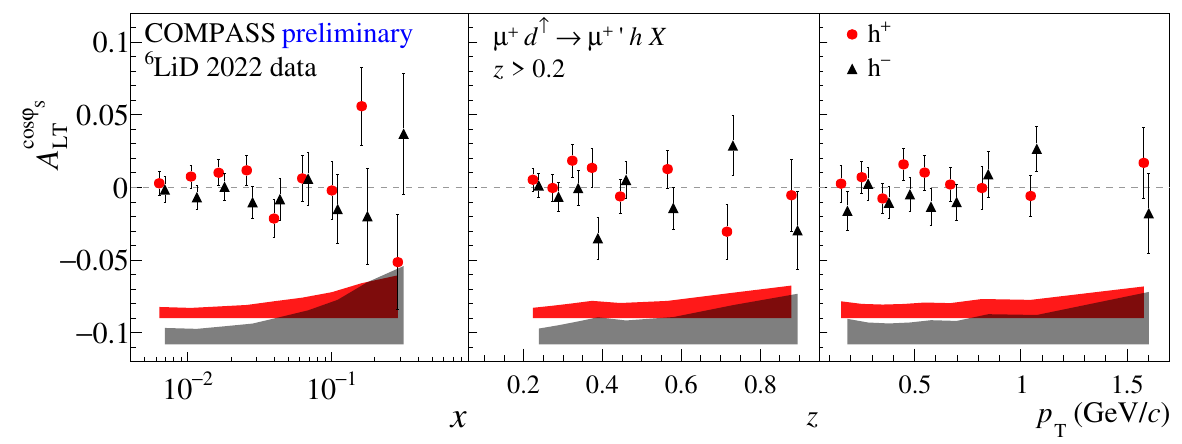}
  \end{minipage}\hfill
  \begin{minipage}[t]{0.49\textwidth}
    \centering
    \includegraphics[width=\linewidth]{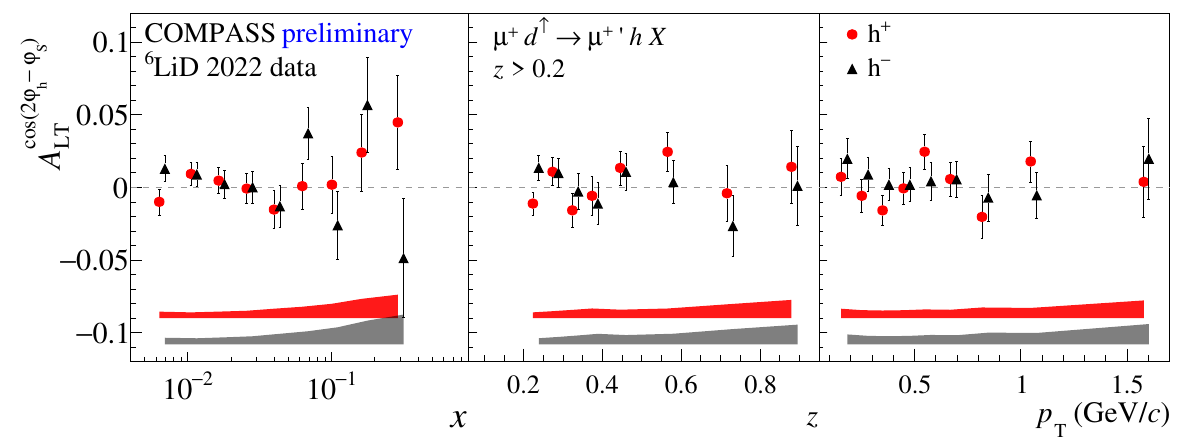}
  \end{minipage}
  \captionof{figure}{Full-statistics 2022 deuteron results for
  $A_{\rm UT}^{\sin\phi_{\rm S}}$ (top left),
  $A_{\rm UT}^{\sin(2\phi_{\rm h}-\phi_{\rm S})}$ (top right),
  $A_{\rm LT}^{\cos\phi_{\rm S}}$ (bottom left), and
  $A_{\rm LT}^{\cos(2\phi_{\rm h}-\phi_{\rm S})}$ (bottom right), as functions of $x$, $z$,
  and $p_{\rm T}$ for positive (red circles) and negative (black triangles)
  hadrons. The error bars represent the statistical uncertainties; the
  red and grey bands mark the systematic uncertainties for positive and
  negative hadrons, respectively. The results are COMPASS preliminary.}
  \label{fig:subleading-twist}
\end{center}

The predominantly small amplitudes, generally compatible with zero, follow a
pattern qualitatively similar to
that observed in earlier COMPASS and HERMES proton measurements, although the
proton $A_{\rm UT}^{\sin\phi_{\rm S}}$ data showed a large-$z$ effect not seen here
~\cite{COMPASS:SixTSA2017,HERMES:2020}. 

\section{Conclusion}

The complete 2022 data set collected with the transversely polarised $^6$LiD
target comprises approximately 114 million selected charged hadrons and
allows the simultaneous extraction of all eight transverse-spin asymmetries
with substantially improved precision over the earlier deuteron
measurements. The full-statistics Collins and Sivers results are consistent
with the published analysis based on approximately 70\% of the 2022 sample.

The Collins amplitudes remain generally small, with an opposite-sign tendency
for positive and negative hadrons at the largest $x$, while the deuteron
Sivers asymmetry is small and compatible with zero. The pretzelosity
asymmetry is compatible with zero, and the Kotzinian--Mulders asymmetry shows
only a possible positive tendency at large $x$. The four subleading-twist
amplitudes are generally small and zero-compatible within uncertainties.
Together, these measurements provide important input for flavour-separated
TMD analyses. 

\section*{Acknowledgements}

This work was supported by the Higher Education and Science Committee of the
Republic of Armenia (research projects No.~23AA-1C023 and
No.~21AG-1C028) and by COST Action CA24159 (SHARP) of COST (European
Cooperation in Science and Technology).

\end{document}